\documentclass{openjournal}

\usepackage{xcolor}
\usepackage{textgreek}
\usepackage[utf8]{inputenc}
\usepackage[english]{babel}
\usepackage{float}
\usepackage[ruled,vlined]{algorithm2e}

\usepackage{hyperref}
\hypersetup{
    unicode, 
    colorlinks=true,
    linkcolor=linkcolor,
    citecolor=linkcolor,
    filecolor=linkcolor,
    urlcolor=linkcolor,
}
\usepackage{color,colortbl}
\definecolor{linkcolor}{rgb}{0.0,0.3,0.5}
\usepackage{tensind}
\tensordelimiter{?}
\DeclareGraphicsExtensions{.bmp,.png,.jpg,.pdf}
\usepackage{verbatim}
\usepackage[normalem]{ulem}
\usepackage{orcidlink}
\usepackage{soul}

\graphicspath{ {./figs/} }

\begin{document}
\title{PDRS : A Linear $\mathcal{O}(N)$ Algorithm for Segmentation of High-Activity Regions in Irregularly Sampled Time Series}

\author{Atal Agrawal\orcidlink{0009-0009-8311-0523}}
\email{atal\_a@ph.iitr.ac.in}
\affiliation{Department of Physics, Indian Institute of Technology Roorkee, Roorkee 247667, India}

\begin{abstract}

Identifying transient high-activity episodes in astronomical time series requires partitioning the data into regions of distinct statistical behavior — a process known as time series segmentation. A widely adopted approach combines Bayesian Blocks, which partition light curves into segments of constant statistical properties, with a subsequent hill-climbing procedure to merge and isolate high-activity regions. However, this combined approach carries an $\mathcal{O}(N^2)$ time complexity, posing a significant scalability challenge for wide-field time-domain surveys like the Zwicky Transient Facility (ZTF) and the upcoming Rubin Observatory (LSST), where individual light curves routinely contain thousands of irregularly sampled observations. We present \textit{Peak-Driven Region Segmentation} (PDRS), a linear-time $\mathcal{O}(N)$ algorithm designed for the rapid extraction of high-activity regions in irregularly sampled data. PDRS seeds candidate regions at statistically significant local maxima and expands them via a gradient-aware multi-source breadth-first search (BFS). To suppress spurious detections, the algorithm employs saddle-point merging and a final median-based filter. Functioning as a computationally efficient pre-processing stage, PDRS isolates candidate transient events for downstream analysis. We demonstrate the efficacy of the algorithm on a sample of quasar light curve from SDSS Stripe~82 and AGN light curves from ZTF Data Release~23, showing that PDRS identifies candidate high-activity regions comparable to those flagged by the Bayesian Blocks approach at substantially reduced computational cost. Because of its domain-agnostic formulation and physically interpretable parameters, PDRS extends beyond astronomy; it is broadly applicable to any irregularly sampled time series exhibiting high-activity episodes, including biomedical signals such as electrocardiograms, seismic ground motion recordings, and industrial sensor monitoring.

\end{abstract}


\maketitle

\section{Introduction}
\label{sec:intro}
Regions of anomalously high activity in a time series signal the occurrence of physical processes that deviate from the quiescent behavior of a source. Identifying and characterizing such regions is therefore essential for understanding the underlying astrophysical mechanisms. Active Galactic Nuclei (AGN) variability is widely studied in this context, as transient high-activity episodes serve as windows into the physics of supermassive black holes and the processes occurring in their vicinity. These include Tidal Disruption Events \cite{10.1093/mnras/stae1036}, Black Hole Mergers \cite{10.1093/mnras/stx1456}, Supernovae \cite{Drake:2011kg}, and Microlensing \cite{10.1093/mnras/stw1963}. Because these events release massive amounts of energy in the form of aperiodic bursts, they must be identified through the direct detection of flux excesses in time-domain light curves. 

Considerable effort has been devoted to detecting this flaring activity in AGN light curves. \cite{10.1093/mnras/stae721} applied Gaussian processes to detect AGN flares in ZTF data. \cite{Meyer2019} combined the Bayesian Blocks algorithm \cite{Scargle2013} with a hill-climbing approach based on the HOP topological grouping method \cite{Eisenstein1998} to characterize the gamma-ray variability of the brightest flat-spectrum radio quasars. This methodology has since been widely adopted: \cite{Kouch2026} applied it to identify flares in blazar light curves from the CRTS, ATLAS, and ZTF surveys, and \cite{He_2025} used it to detect flares in AGN light curves from ZTF Data Release~23. 

While the Bayesian Blocks algorithm is a powerful dynamic programming tool that partitions a time series into segments of statistically constant flux, its $\mathcal{O}(N^2)$ time complexity \cite{Scargle2013} creates a significant computational bottleneck. This quadratic scaling is a notable limitation for current surveys like the Zwicky Transient Facility (ZTF) and upcoming facilities such as the Rubin Observatory Legacy Survey of Space and Time (LSST), where individual light curves routinely contain thousands of observations. In this work, we present \textit{Peak-Driven Region Segmentation} (PDRS), a linear-time $\mathcal{O}(N)$ algorithm (Algorithm~\ref{alg:pdrs}) designed as a scalable alternative to the \cite{Meyer2019} approach for identifying high-activity regions in large-scale time series datasets. 

The remainder of this paper is structured as follows. Section~\ref{sec:algorithm} describes the core logic and implementation of the PDRS algorithm. Section~\ref{sec:discussion} presents our results and discussion, Section~\ref{sec:conclusions} summarizes our conclusions and plots of PDRS applied to ZTF DR23 and SDSS stripe 82 quasar data are present in Appendix~\ref{app:additional}.
\section{PDRS Algorithm}
\label{sec:algorithm}

PDRS algorithm operates in four sequential stages: (1) peak identification, (2) gradient-aware region expansion, (3) saddle-point merging, and (4) median-based filtering. Each stage is described below.
\begin{algorithm}[]
\caption{Peak-Driven Region Segmentation (PDRS)}
\label{alg:pdrs}
\SetAlgoLined
\KwIn{Time series $(t, f)$, parameters $\sigma_{\mathrm{thresh}}, r_{\mathrm{saddle}}, N_{\mathrm{min}}, w_{\mathrm{smooth}}, \sigma_{\mathrm{region}}, \Delta t_{\mathrm{max}}$}
\KwOut{Set of candidate regions $\mathcal{R}$}
\tcp{Stage 1: Peak Identification}
Compute $\mu \leftarrow \mathrm{median}(f)$ and $\sigma \leftarrow \mathrm{std}(f)$\;
Compute smoothed gradient $g \leftarrow \textsc{LocalGradient}(t, f, w_{\mathrm{smooth}})$\;
Identify local maxima: candidate peak at index $i$ if $f_i > f_{i-1}$ and $f_i > f_{i+1}$\;
Also check endpoints: include index $0$ if $f_0 > f_1$, and index $N-1$ if $f_{N-1} > f_{N-2}$\;
Retain only peaks where $f_i > \mu + \sigma_{\mathrm{thresh}} \cdot \sigma$; collect in $\mathcal{P}$\;
\If{$\mathcal{P}$ is empty}{\Return $\emptyset$\;}
\tcp{Stage 2: Gradient-Aware Multi-Source BFS Expansion}
Mark all points as unassigned\;
For each peak $p \in \mathcal{P}$: initialise left and right frontier at $p$, mark $p$ as assigned to $p$\;
\tcp{All frontiers expand simultaneously; a point is claimed by the first frontier to reach it}
\While{any frontier is active}{
    \tcp{Each peak advances its frontier by one step per iteration}
    \ForEach{peak $p \in \mathcal{P}$ with active frontier}{
        $\mathrm{expanded} \leftarrow \mathrm{False}$\;
        \tcp{Expand left: only if neighbor is still unassigned}
        \If{left neighbor is unassigned, flux $\geq \mu$, and time gap $\leq \Delta t_{\mathrm{max}}$}{
            \uIf{left neighbor flux $<$ current left frontier flux}{
                Assign left neighbor to $p$; advance left frontier; $\mathrm{expanded} \leftarrow \mathrm{True}$\;
            }
            \ElseIf{gradient at left frontier $\geq 0$}{
                Assign left neighbor to $p$; advance left frontier; $\mathrm{expanded} \leftarrow \mathrm{True}$\;
            }
        }
        \tcp{Expand right: only if neighbor is still unassigned}
        \If{right neighbor is unassigned, flux $\geq \mu$, and time gap $\leq \Delta t_{\mathrm{max}}$}{
            \uIf{right neighbor flux $<$ current right frontier flux}{
                Assign right neighbor to $p$; advance right frontier; $\mathrm{expanded} \leftarrow \mathrm{True}$\;
            }
            \ElseIf{gradient at right frontier $\leq 0$}{
                Assign right neighbor to $p$; advance right frontier; $\mathrm{expanded} \leftarrow \mathrm{True}$\;
            }
        }
        \If{not $\mathrm{expanded}$}{Deactivate frontier of $p$\;}
    }
}
Collect clusters: for each $p \in \mathcal{P}$, retain cluster if it contains $\geq N_{\mathrm{min}}$ points\;
\If{no clusters remain}{\Return $\emptyset$\;}

\tcp{Stage 3: Saddle-Point Merging}
Initialise merged list with first cluster; set running peak $\leftarrow$ peak flux of first cluster\;
\ForEach{subsequent cluster}{
    Compute temporal gap $\leftarrow t_{\mathrm{start}}^{\mathrm{curr}} - t_{\mathrm{end}}^{\mathrm{prev}}$\;
    \uIf{temporal gap $> \Delta t_{\mathrm{max}}$}{
        \tcp{Large data void — always separate}
        Add current cluster to merged list\;
        Update running peak $\leftarrow$ current peak flux\;
    }
    \uElseIf{current cluster start $\leq$ previous cluster end $+ 2$}{
        \tcp{Overlap or proximity: merge unconditionally}
        Extend previous cluster end to current cluster end\;
        Update running peak $\leftarrow \max($running peak, current peak flux$)$\;
    }
    \Else{
        \tcp{True saddle exists: compute minimum flux strictly between cluster boundaries}
        Compute saddle flux $\leftarrow \min(f$ strictly between previous cluster end and current cluster start$)$\;
        \If{$(\mathrm{saddle\ flux} - \mu) > r_{\mathrm{saddle}} \cdot (\min(\mathrm{running\ peak},\ \mathrm{current\ peak\ flux}) - \mu)$}{
            Extend previous cluster end to current cluster end\;
            Update running peak $\leftarrow \max($running peak, current peak flux$)$\;
        }
        \Else{
            Add current cluster to merged list\;
            Update running peak $\leftarrow$ current peak flux\;
        }
    }
}

\tcp{Stage 4: Median-Based Filtering}
$\mathcal{R} \leftarrow \emptyset$\;
\ForEach{merged cluster}{
    \If{median flux of cluster $\geq \mu + \sigma_{\mathrm{region}} \cdot \sigma$}{
        Add $(t_{\mathrm{start}},\ t_{\mathrm{end}},\ \mathrm{peak\ flux})$ to $\mathcal{R}$\;
    }
}
\Return $\mathcal{R}$\;
\end{algorithm}

\subsection{Preprocessing: Temporal Binning}
Raw photometric light curves from time-domain surveys typically contain 
multiple observations per night with varying cadence and photometric 
uncertainties. Prior to binning, observations with non-zero photometric 
flags are excluded, as major time-domain surveys such as ZTF apply 
quality flagging at the pipeline level. The light curve is then binned 
into fixed-width temporal bins of size $\Delta t_{\mathrm{bin}}$ days 
using inverse-variance weighted averaging, which suppresses intra-bin 
noise while preserving the overall flux structure. The binned flux and 
its uncertainty are propagated through to the detection stage. This 
binning step is $\mathcal{O}(N)$ and does not affect the overall 
complexity of the algorithm.

\subsection{Stage 1: Peak Identification}
All detection and segmentation operations are performed in flux space, 
as flux is a linear measure of source brightness and better preserves 
the statistical properties required for threshold-based detection. 
The final output is converted back to magnitudes for visualization purposes.
The algorithm begins by computing the global baseline of the light curve. The 
median flux $\mu$ and standard deviation $\sigma$ are computed over all $N$ 
data points. The median is preferred over the mean as it is robust to the 
presence of flaring activity, which would otherwise bias the baseline upward. 
A smoothed local gradient $g$ is then estimated at each point using a 
sliding window linear regression of size $w_{\mathrm{smooth}}$, computed in 
$\mathcal{O}(N)$ via prefix sums of the relevant statistical moments 
($t$, $f$, $t^2$, and $t \cdot f$). This gradient 
is used in the subsequent expansion stage to handle non-monotonic flux 
variations near flare boundaries.

Local maxima are identified as points where the flux exceeds both its immediate 
neighbors. A candidate peak must additionally satisfy $f_i > \mu + 
\sigma_{\mathrm{thresh}} \cdot \sigma$, ensuring that only statistically 
significant flux excesses seed the expansion. Points that do not satisfy this 
condition are discarded, preventing noise fluctuations near the baseline from 
initiating spurious regions.

\subsection{Stage 2: Gradient-Aware Multi-Source BFS Expansion}
Each candidate peak $p \in \mathcal{P}$ is assigned a frontier $[l_p, r_p]$, 
initialised at $[p, p]$. All frontiers expand simultaneously in a multi-source 
breadth-first search, meaning that each unassigned point can be claimed by at 
most one peak — whichever frontier reaches it first. This prevents overlapping 
regions and ensures that boundaries between adjacent flares are drawn at the 
natural flux minimum between them.

At each step, the left frontier attempts to expand to $l_p - 1$ and the right 
frontier to $r_p + 1$. Expansion is permitted only if four conditions are 
simultaneously satisfied: (1) the candidate point is unassigned, (2) its flux 
is at or above the global median $\mu$, (3) the temporal gap to the frontier 
does not exceed $\Delta t_{\mathrm{max}}$, and (4) the flux trend supports 
inclusion. For condition (4), strict monotonicity is checked first — if the 
flux at the candidate point is lower than at the current frontier, expansion 
proceeds unconditionally. If monotonicity is broken (i.e., a local dip or 
plateau is encountered), the algorithm falls back to the pre-computed gradient: 
left expansion requires $g_{l_p} \geq 0$ and right expansion requires 
$g_{r_p} \leq 0$, ensuring the smoothed trend still supports the direction of 
expansion. A frontier is deactivated once neither left nor right expansion is 
possible. After all frontiers are deactivated, clusters with fewer than 
$N_{\mathrm{min}}$ points are discarded.

\subsection{Stage 3: Saddle-Point Merging}
A single physical flare may produce multiple candidate peaks if its light 
curve contains minor sub-structure or noise fluctuations near the top. To 
consolidate such cases, adjacent clusters are considered for merging under 
three conditions. First, if the temporal gap between the end of one cluster 
and the start of the next exceeds $\Delta t_{\mathrm{max}}$, the clusters 
are always kept separate, as no meaningful saddle flux can be defined across 
a data void. Second, if two clusters overlap or are separated by fewer than 
two data points, they are merged unconditionally, as the gap is too small to 
constitute a meaningful valley. Third, if the saddle flux 
$f_{\mathrm{saddle}}$ --- defined as the minimum flux strictly between 
the end of one cluster and the start of the next, excluding the boundary 
points themselves --- satisfies $(f_{\mathrm{saddle}} - \mu) > 
r_{\mathrm{saddle}} \cdot (\min(f_{\mathrm{curr}}, f_{\mathrm{peak},i}) 
- \mu)$, the valley is considered too shallow to warrant a separation 
and the clusters are merged. Here $f_{\mathrm{curr}}$ is the peak flux 
of the current merged cluster (i.e., the strongest peak accumulated so 
far in the merge chain), and $f_{\mathrm{peak},i}$ is the peak flux of 
the incoming cluster. The parameter $r_{\mathrm{saddle}}$ controls the 
minimum fractional elevation of the inter-cluster saddle above the global 
median $\mu$, relative to the weaker of the two peak elevations above 
$\mu$. A lower value of $r_{\mathrm{saddle}}$ permits merging when the 
saddle is only weakly elevated above $\mu$, while a higher value requires 
the saddle to be more substantially elevated before a merge is triggered.

\subsection{Stage 4: Median-Based Filtering}
A candidate region may survive the BFS expansion and minimum cluster 
size requirement while still being driven by a sparse collection of 
randomly scattered high-flux points rather than a genuinely sustained 
flux excess. To suppress such spurious detections, a final gate requires 
that the median flux of the entire candidate region satisfies 
$\mathrm{median}(f_{s:e}) \geq \mu + \sigma_{\mathrm{region}} \cdot \sigma$. 
Since the median is robust to isolated high values, a region populated 
by randomly scattered flux spikes will fail this criterion even if its 
peak and cluster size are sufficient. Regions that fail this criterion 
are discarded. The surviving regions constitute the final output 
$\mathcal{R}$, each characterised by a start time $t_s$, end time $t_e$, 
and peak flux $\max(f_{s:e})$.

\subsection{Computational Complexity}
Each stage of PDRS operates in linear time. The preprocessing and peak 
identification in Stage~1 require a single pass over the data, $\mathcal{O}(N)$. 
The gradient estimation uses prefix sums and is likewise $\mathcal{O}(N)$. 
The multi-source BFS in Stage~2 assigns each point at most once across all 
frontiers, giving $\mathcal{O}(N)$ total. The saddle-point merging in 
Stage~3 and median filtering in Stage~4 operate over the set of clusters, 
which is bounded by the number of peaks and therefore $\mathcal{O}(N)$ in 
the worst case. The overall time complexity of PDRS is therefore 
$\mathcal{O}(N)$, in contrast to the $\mathcal{O}(N^2)$ complexity of the 
Bayesian Blocks algorithm.

\begin{table}[]
    \centering
    \renewcommand{\arraystretch}{1.4}
    \begin{tabular}{ c p{0.75\linewidth} }  
        \hline \hline     
        \textbf{Parameter} & \textbf{Description} \\
        \hline      
        $\sigma_{\mathrm{thresh}}$ & Minimum peak significance above the global median (in units of $\sigma$). \\
        $r_{\mathrm{saddle}}$      & Minimum fractional elevation of the inter-cluster saddle above the global median $\mu$, relative to the weaker of the two peak elevations above $\mu$. \\
        $N_{\mathrm{min}}$         & Minimum number of data points required to define a valid region. \\
        $w_{\mathrm{smooth}}$      & Window size for the local gradient estimation. \\
        $\sigma_{\mathrm{region}}$ & Minimum median flux of a candidate region above the global median (in units of $\sigma$). \\
        $\Delta t_{\mathrm{max}}$  & Maximum temporal gap (days) permitted during region expansion to prevent crossing observing gaps. \\
        \hline \hline  
    \end{tabular}  
    \caption{Free parameters of the PDRS algorithm and their roles.}
    \label{table:pdrs_params}
\end{table}
\vspace{8mm}

\section{Results and Discussion}
\label{sec:discussion}

\subsection{Free Parameters}
\label{sec:parameters}
The algorithm has six free parameters, summarised in Table~\ref{table:pdrs_params}. 
$\sigma_{\mathrm{thresh}}$ controls the minimum significance of a candidate peak 
above the global baseline. It must be tuned to the contrast between flaring 
activity and the quiescent level of the source. For ZTF Data Release~23 AGN 
light curves we adopt $\sigma_{\mathrm{thresh}} = 2$, while for SDSS Stripe~82 quasar light curves a lower value of 
$\sigma_{\mathrm{thresh}} = 1$ is used, as quasars at cosmological distances 
appear intrinsically faint, causing their flaring activity to produce only 
modest flux excesses above the quiescent baseline that would be missed at a 
higher threshold.
$N_{\mathrm{min}}$ sets the minimum number of data points required to constitute 
a valid candidate region, ensuring that isolated photometric defects or 
single-point spikes are not retained as flare candidates.
$\Delta t_{\mathrm{max}}$ prevents the BFS expansion from bridging seasonal 
observing gaps. For ZTF, where seasonal gaps are of the order of months, we 
set $\Delta t_{\mathrm{max}} = 60$ days. For SDSS Stripe~82, which has longer 
inter-season gaps, we adopt $\Delta t_{\mathrm{max}} = 200$ days.
$\sigma_{\mathrm{region}}$ acts as a final gate on the merged candidate regions, 
requiring that the median flux of the entire region exceeds the baseline by at 
least $\sigma_{\mathrm{region}} \cdot \sigma$. This ensures that a region is not 
retained on the basis of a few isolated spikes but represents a genuinely 
sustained flux excess. This filter is applied after saddle-point merging, by 
which point the region boundary already reflects the physical structure of the 
flare. We set $\sigma_{\mathrm{region}} = 0.5$ for both ZTF data and SDSS 
Stripe~82 data.
$w_{\mathrm{smooth}}$ controls the window size for the local gradient estimation 
used during BFS expansion. It should be chosen relative to the cadence of the 
data, as a larger window smooths over short-timescale flux fluctuations during 
the rise and decay phases of a flare, allowing the expansion to navigate minor 
dips without prematurely terminating.
$r_{\mathrm{saddle}}$ governs the saddle-point merging criterion. Two adjacent clusters are merged if the flux at the inter-cluster 
saddle is sufficiently elevated above the global median $\mu$, as 
quantified by $r_{\mathrm{saddle}}$, relative to the weaker of the 
two peak elevations above $\mu$. This merging step is conceptually 
inspired by the basin-merging logic of the HOP algorithm 
\cite{Eisenstein1998}, adapted here to operate on the elevation 
above the global baseline rather than absolute valley depth.

\subsection{Comparison with Bayesian Blocks}
\label{sec:results}

We apply PDRS to flaring AGN light curves from ZTF Data Release~23, drawn from the AGNFRC catalog \cite{He_2025}, which provides a sample of high-confidence flare labels. For comparison, we process the same light curves using the combined Bayesian Blocks and hill-climbing methodology, adopting the implementation described by \cite{Meyer2019} and \cite{He_2025}. For the PDRS configuration, we fix the algorithmic parameters as follows: peak significance threshold $\sigma_{\mathrm{thresh}} = 2$, saddle-point depth ratio $r_{\mathrm{saddle}} = 0.2$, minimum cluster size $N_{\mathrm{min}} = 3$, gradient smoothing window $w_{\mathrm{smooth}} = 7$, region median threshold $\sigma_{\mathrm{region}} = 0.5$, and maximum temporal gap $\Delta t_{\mathrm{max}} = 60$\,days.

\begin{figure}[ht!]
    \centering
    \includegraphics[width=\linewidth]{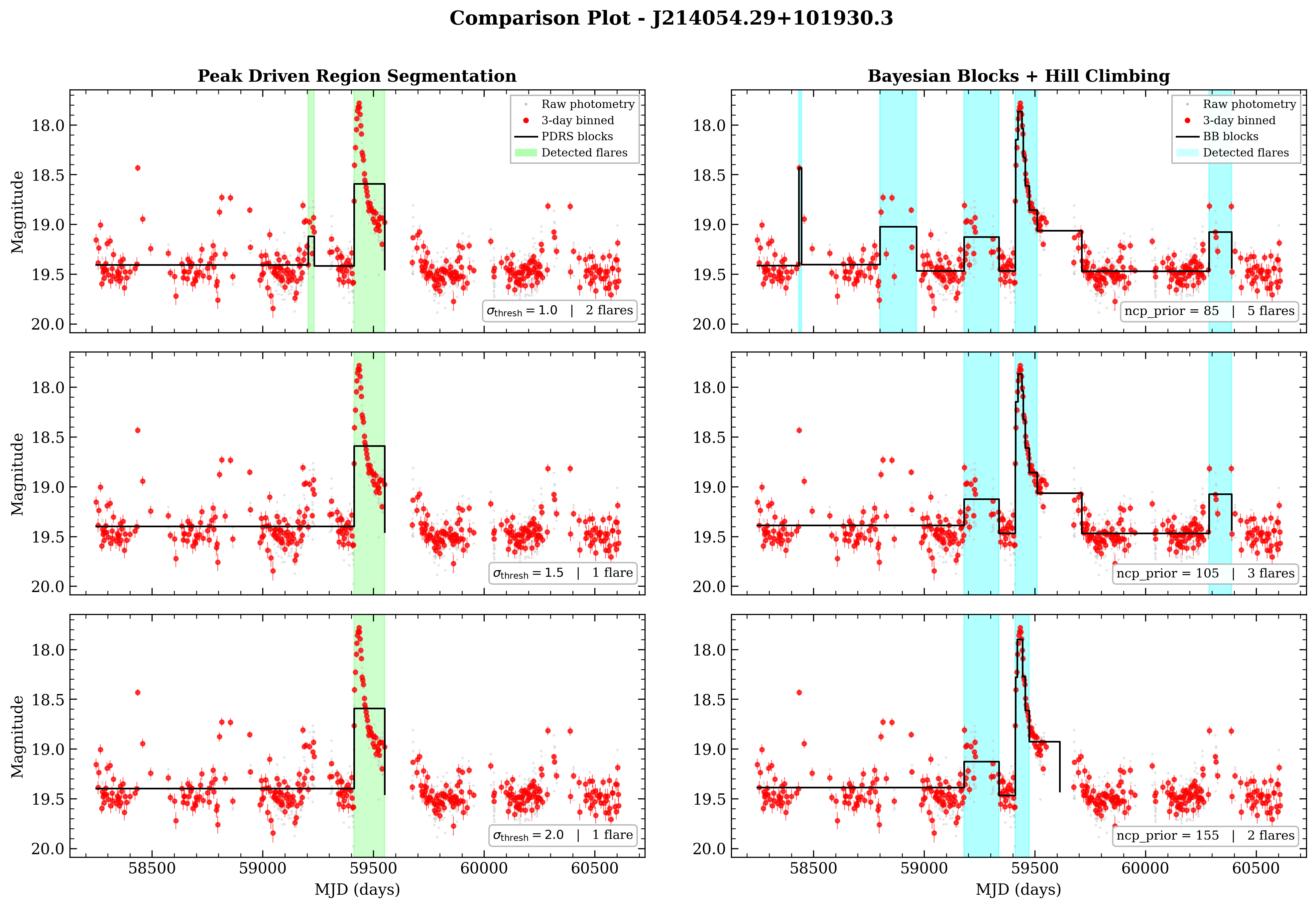}
    \caption{Comparison of PDRS (left) and Bayesian Blocks combined with 
hill-climbing (right) applied to the ZTF light curve of 
J214054.29+101930.3, an AGN from the AGNFRC catalog \cite{He_2025}. 
Raw photometry (grey points) is binned into 3-day bins (red points) 
prior to detection. Each row corresponds to a different parameter 
setting: $\sigma_{\mathrm{thresh}} = 1.0, 1.5, 2.0$ for PDRS and 
\texttt{ncp\_prior} $= 85, 105, 155$ for Bayesian Blocks. PDRS 
detects 2 regions at $\sigma_{\mathrm{thresh}} = 1.0$ and converges 
to a single candidate high-activity region at higher thresholds, 
capturing the prominent flux excess near MJD~59500 including its 
rise and decay. The Bayesian Blocks approach detects between 2 and 5 
regions depending on \texttt{ncp\_prior}, flagging several noise 
fluctuations as candidate flares at lower prior values.}
    \label{fig:comparison}
\end{figure}

As demonstrated in Figure~\ref{fig:comparison}, PDRS yields highly 
stable segmentations across a range of peak significance thresholds 
($\sigma_{\mathrm{thresh}}$). At $\sigma_{\mathrm{thresh}} = 1.0$, 
PDRS detects 2 candidate regions, converging to a single 
high-activity region at higher thresholds, accurately encompassing 
both the rise and decay phases of the event while ignoring stochastic 
background fluctuations. In contrast, the Bayesian Blocks approach 
exhibits high sensitivity to its penalty parameter (\texttt{ncp\_prior}). 
Even at elevated penalties, the dynamic programming method is prone to 
over-segmenting the baseline noise, resulting in a highly variable 
number of spurious detected regions depending on the chosen prior. 

Additional demonstrations of the PDRS framework, including its application to a multi-year SDSS Stripe~82 quasar light curve containing a previously identified flare \cite{Agrawal2026} and further example from the AGNFRC catalog, are provided in Appendix~\ref{app:additional}.

\vspace{5mm}

\section{Conclusions}
\label{sec:conclusions}

In this work, we present \textit{Peak-Driven Region Segmentation} 
(PDRS), an $\mathcal{O}(N)$ algorithm designed to efficiently isolate 
transient high-activity regions in irregularly sampled time series for 
downstream physical analysis. PDRS serves as a scalable alternative to 
the widely adopted Bayesian Blocks and hill-climbing pipeline, which 
carries an $\mathcal{O}(N^2)$ computational cost and can over-segment 
quiescent background noise depending on the choice of penalty parameter. 
PDRS instead adopts a targeted, peak-first approach, seeding candidate 
regions at statistically significant local maxima and expanding them via 
a gradient-aware multi-source breadth-first search, extracting flaring 
events in linear time.

A trade-off of the PDRS framework is its reliance on a set of 
user-defined free parameters. However, unlike abstract algorithmic 
penalties, these parameters are physically interpretable. The inclusion 
of morphological constraints---specifically the minimum cluster size 
($N_{\mathrm{min}}$), the region median threshold ($\sigma_{\mathrm{region}}$), 
and the maximum temporal gap ($\Delta t_{\mathrm{max}}$)---provides 
researchers with granular control to suppress false-positive detections. 
The computational efficiency of PDRS makes it well-suited as a rapid 
pre-processing stage for the massive data volumes anticipated from 
upcoming wide-field time-domain surveys, while its domain-agnostic 
architecture ensures broad applicability to any discipline analyzing 
stochastic, bursty signals.

\section*{Acknowledgments}
We acknowledge \cite{Masci2019ZTF} for the ZTF DR23 data and 
\cite{2010ApJ...721.1014M} for the SDSS Stripe~82 quasar data 
used to demonstrate the algorithm.

Based on observations obtained with the Samuel Oschin Telescope 48-inch and the 60-inch Telescope at the Palomar
Observatory as part of the Zwicky Transient Facility project. ZTF is supported by the National Science Foundation under Grants
No. AST-1440341 and AST-2034437 and a collaboration including current partners Caltech, IPAC, the Oskar Klein Center at
Stockholm University, the University of Maryland, University of California, Berkeley , the University of Wisconsin at Milwaukee,
University of Warwick, Ruhr University, Cornell University, Northwestern University and Drexel University. Operations are
conducted by COO, IPAC, and UW.
Funding for the Sloan Digital Sky Survey V has been provided by the Alfred P. Sloan Foundation, the Heising-Simons Foundation, the National Science Foundation, and the Participating Institutions. SDSS acknowledges support and resources from the Center for High-Performance Computing at the University of Utah. SDSS telescopes are located at Apache Point Observatory, funded by the Astrophysical Research Consortium and operated by New Mexico State University, and at Las Campanas Observatory, operated by the Carnegie Institution for Science. The SDSS web site is \url{www.sdss.org}.

SDSS is managed by the Astrophysical Research Consortium for the Participating Institutions of the SDSS Collaboration, including the Carnegie Institution for Science, Chilean National Time Allocation Committee (CNTAC) ratified researchers, Caltech, the Gotham Participation Group, Harvard University, Heidelberg University, The Flatiron Institute, The Johns Hopkins University, L'Ecole polytechnique f\'{e}d\'{e}rale de Lausanne (EPFL), Leibniz-Institut f\"{u}r Astrophysik Potsdam (AIP), Max-Planck-Institut f\"{u}r Astronomie (MPIA Heidelberg), Max-Planck-Institut f\"{u}r Extraterrestrische Physik (MPE), Nanjing University, National Astronomical Observatories of China (NAOC), New Mexico State University, The Ohio State University, Pennsylvania State University, Smithsonian Astrophysical Observatory, Space Telescope Science Institute (STScI), the Stellar Astrophysics Participation Group, Universidad Nacional Aut\'{o}noma de M\'{e}xico, University of Arizona, University of Colorado Boulder, University of Illinois at Urbana-Champaign, University of Toronto, University of Utah, University of Virginia, Yale University, and Yunnan University.  

\noindent\textit{Software:} NumPy \cite{numpy},
Matplotlib \cite{matplotlib},
Pandas \cite{pandas}

\section*{Data and Software Availability}
The Python implementation of the Peak-Driven Region Segmentation (PDRS) algorithm presented in this work, along with the script necessary to reproduce the figures, is open-source and publicly available on GitHub at \url{https://github.com/zerozole/Peak_Driven_Region_Segmentation}.

\bibliographystyle{apsrev4-1}

\bibliography{oja_template}

\appendix
\section{Application of PDRS to ZTF DR23 and SDSS Stripe~82 quasar Light Curves}
\label{app:additional}

In this appendix, we present supplementary examples of astronomical light curves segmented using the PDRS algorithm. For the SDSS Stripe~82 quasar light curve, we adopted the foundational parameters defined in Section~\ref{sec:parameters}, supplemented by the specific configuration $r_{\mathrm{saddle}} = 0.2$, $N_{\mathrm{min}} = 3$, and $w_{\mathrm{smooth}} = 7$. For the ZTF Data Release~23 AGN light curve, we utilized the identical suite of free parameters established in Section~\ref{sec:results}.

\clearpage

\begin{figure}[t]
    \centering
    \includegraphics[width=0.8\linewidth]{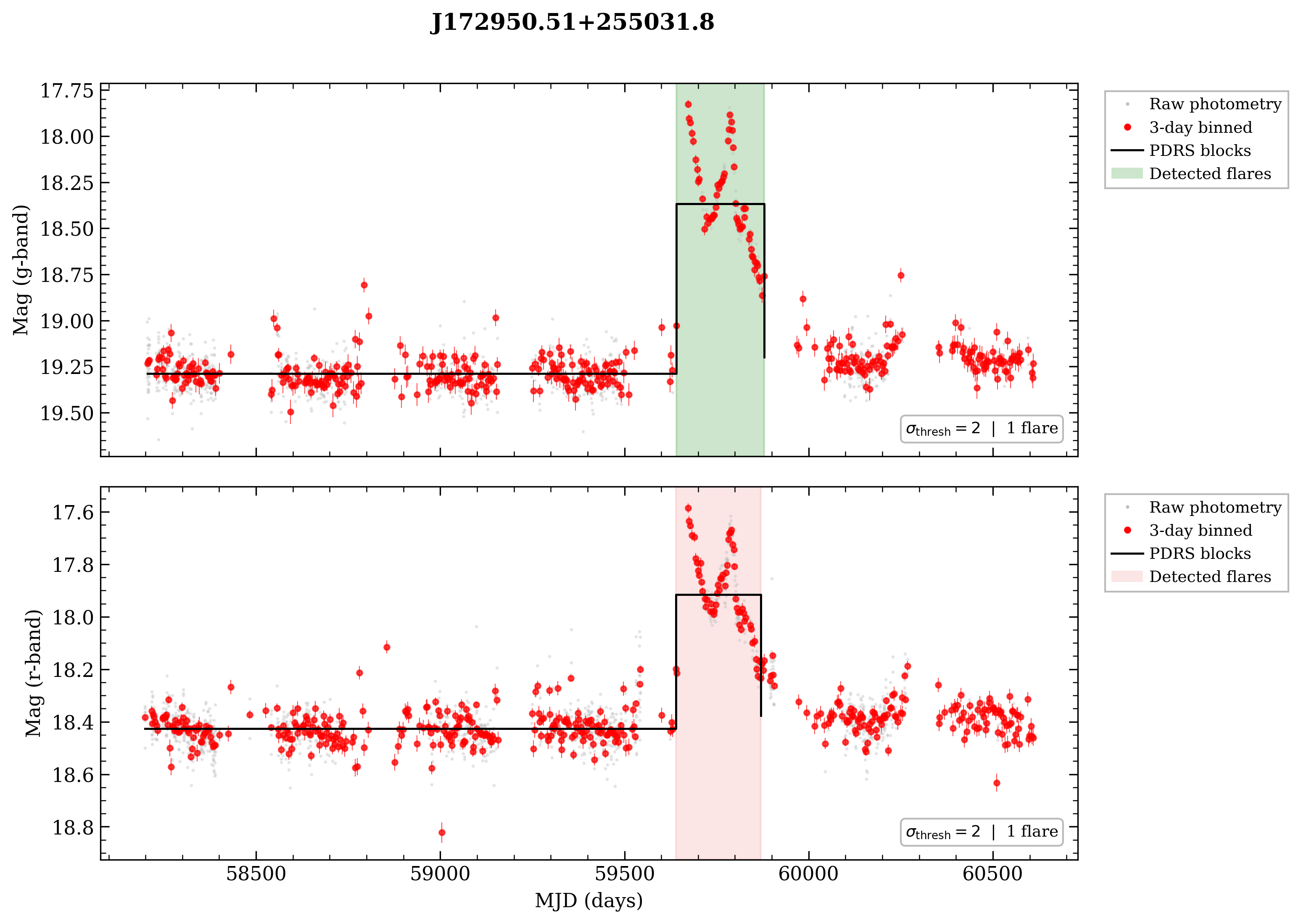}
    \caption{Application of the PDRS algorithm to the multi-band light curve of 
J172950.51+255031.8, an AGN from the AGNFRC catalog \cite{He_2025}. 
The top and bottom panels display the $g$-band and $r$-band photometry, 
respectively. Raw observations are shown in grey, overlaid with 3-day 
binned data (red). The solid black line illustrates the baseline and 
elevated states defined by the PDRS segmentation blocks. Operating 
independently on each band with a peak significance threshold of 
$\sigma_{\mathrm{thresh}} = 2$, the algorithm successfully isolates a 
broadly contemporaneous flaring event (shaded regions), demonstrating 
its robustness across different optical filters.}
    \label{fig:ztf}
\end{figure}

\begin{figure}[t]
    \centering
    \includegraphics[width=0.8\linewidth]{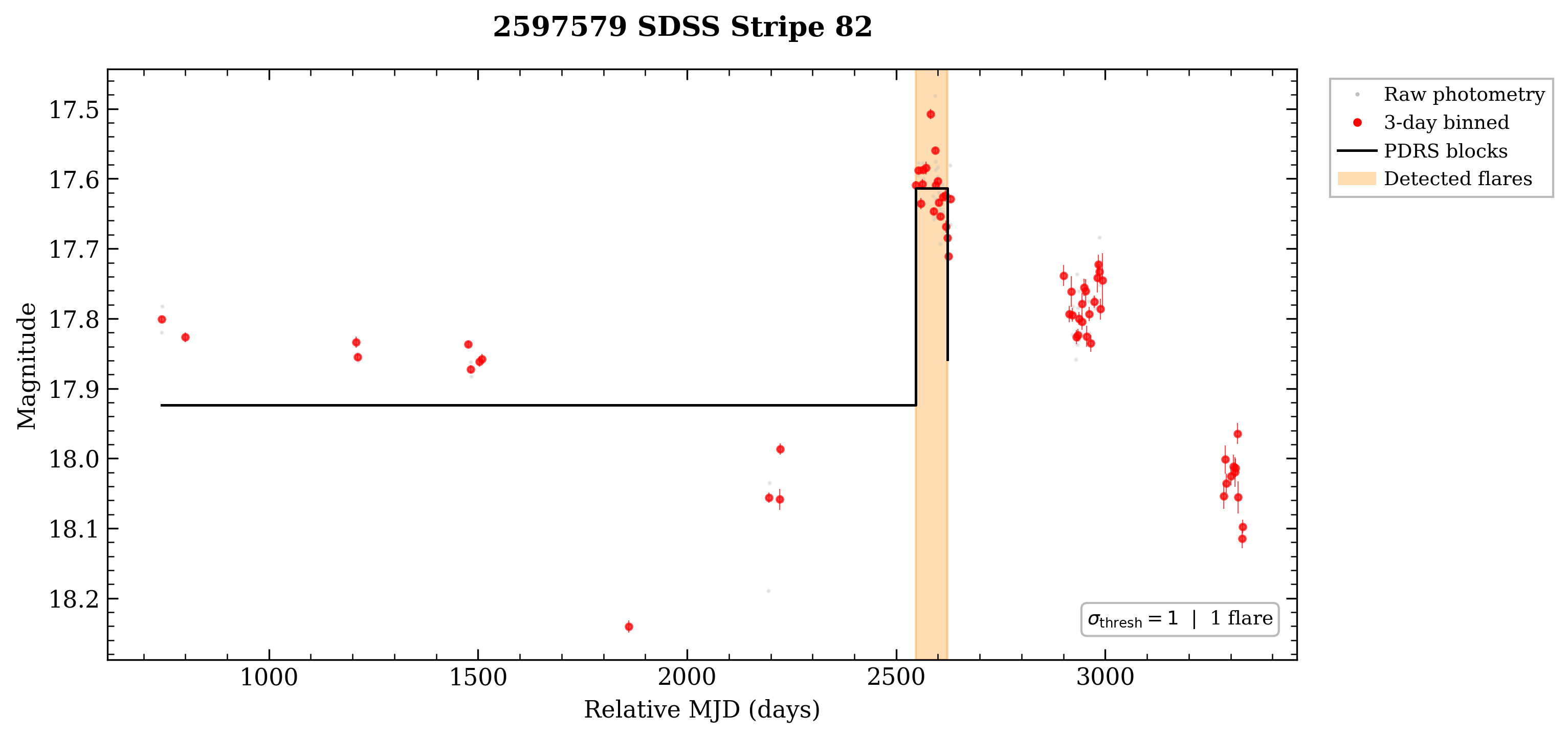}
    \caption{Application of the PDRS algorithm to the light curve of the SDSS 
Stripe~82 quasar 2597579 \cite{Agrawal2026}. The 3-day binned data 
are shown in red; the sparse cadence of Stripe~82 renders the raw and 
binned photometry nearly indistinguishable at this scale. The solid 
black line represents the baseline and elevated states defined by the 
PDRS segmentation blocks. Operating with a peak significance threshold 
of $\sigma_{\mathrm{thresh}} = 1$, the algorithm successfully identifies 
and isolates a distinct flaring episode (shaded orange region) despite 
the prominent seasonal gaps and long-term variability characteristic of 
sparsely sampled historical survey data. The narrow extent of the 
detected region reflects the sparse cadence of Stripe~82 rather than 
an intrinsically short flare duration.}
    \label{fig:sdss}
\end{figure}

\end{document}